\documentclass[aps,prd,preprint,preprintnumbers,unsortedaddress,superscriptaddress,showpacs,nofootinbib]{revtex4-1}
\usepackage{graphicx,color}
\usepackage{relsize}
\usepackage{slashed}
\usepackage{ulem}
\usepackage{tabu}
\usepackage{bm}

\usepackage{amsmath}
\usepackage{amssymb}
\usepackage{amsthm}
\usepackage{mathrsfs}
\usepackage{graphicx}
\usepackage{epstopdf}
\usepackage{fancyhdr}
\usepackage{array}
\usepackage[all]{xy}
\usepackage{eufrak}
\usepackage{euscript}
\usepackage{enumerate}
\usepackage{slashed}
\usepackage{hyperref}
\usepackage{subfigure}
\usepackage{epstopdf}

\hypersetup{pdftex,colorlinks=true,linkcolor=blue,citecolor=blue,menucolor=black,urlcolor=blue,filecolor=blue}

\begin{document}

\title{$\boldsymbol{\Omega_b^- \to (\Xi_c^+ \, K^-) \, \pi^-}$ and the $\boldsymbol{\Omega_c}$ states}

\author{V.~R.~Debastiani}
\email{vinicius.rodrigues@ific.uv.es}
\affiliation{Departamento de F\'{i}sica Te\'{o}rica and IFIC, Centro Mixto Universidad de Valencia - CSIC,
Institutos de Investigaci\'{o}n de Paterna, Aptdo. 22085, 46071 Valencia, Spain.}

\author{J.~M.~Dias}
\email{jdias@if.usp.br}
\affiliation{Departamento de F\'{i}sica Te\'{o}rica and IFIC, Centro Mixto Universidad de Valencia - CSIC,
Institutos de Investigaci\'{o}n de Paterna, Aptdo. 22085, 46071 Valencia, Spain.}
\affiliation{Instituto de F\'{i}sica, Universidade de S\~{a}o Paulo, Rua do Mat\~{a}o, 1371, Butant\~{a}, CEP 05508-090, S\~{a}o Paulo, S\~{a}o Paulo, Brazil}

\author{Wei-Hong Liang}
\email{liangwh@gxnu.edu.cn}
\affiliation{Department of Physics, Guangxi Normal University, Guilin 541004, China}

\author{E.~Oset}
\email{eulogio.oset@ific.uv.es}
\affiliation{Departamento de F\'{i}sica Te\'{o}rica and IFIC, Centro Mixto Universidad de Valencia - CSIC,
Institutos de Investigaci\'{o}n de Paterna, Aptdo. 22085, 46071 Valencia, Spain.}

\date{\today}

\begin{abstract}
We study the weak decay $\Omega_b^- \to (\Xi_c^+ \, K^-) \, \pi^-$, in view of the narrow $\Omega_c$ states recently measured by the LHCb collaboration and later confirmed by the Belle collaboration. The $\Omega_c(3050)$ and $\Omega_c(3090)$ are described as meson-baryon molecular states, using an extension of the local hidden gauge approach in coupled channels. We investigate the $\Xi D$, $\Xi_c \bar K$ and $\Xi_c^\prime \bar K$ invariant mass distributions making predictions that could be confronted with future experiments, providing useful information that could help determine the quantum numbers and nature of these states.
\end{abstract}

\maketitle

\section{Introduction}

The recent discovery of five narrow $\Omega_c$ states by the LHCb collaboration \cite{OmegacLHCb} in $pp$ collisions, also recently confirmed by the Belle collaboration \cite{OmegacBelle} in $e^+e^-$ collisions, motivated an increasing amount of theoretical work with different proposals for their structure. In particular, the correct assignment of quantum numbers $J^P$ remains an open question and it could be the key to understand the nature of these states.

Predictions using quark models for such states and related ones were done in Refs.~\cite{ebert,roberts,garcilazo,migura,ebert2,valcarce,shah,vijande,yoshida,cheng,cheng2,chiladze,chiralquark}.
Quark models mostly propose a diquark-quark structure $(ss)c$, where it is assumed that the diquark $(ss)$ is in the ground state $1S$ bound by the attractive antitriplet color structure $\bar{3}$; and in this case the identical flavor content of the diquark implies that it has spin $S_{ss}=1$, which then can combine with the spin of the $c$ quark $S_c$ and with orbital momentum $L$ of the diquark $(ss)$ relative to the $c$ quark. One of the typical assignments in the literature consists in the five $1P$ possibilities of combining angular momentum when $L=1$, which yields two states with $J^P=1/2^-$, two with $3/2^-$ and one with $5/2^-$, that split due to spin-dependent forces and could correspond to the five narrow $\Omega_c$ states: $\Omega_c(3000)$, $\Omega_c(3050)$, $\Omega_c(3066)$, $\Omega_c(3090)$ and $\Omega_c(3119)$, in the same order of increasing $J$. Alternative assignments also include the possibility of a radial excitation of the diquark $(ss)$ relative to the $c$ quark, where the last two $\Omega_c$ states would be $2S$ excitations with quantum numbers $J^P=1/2^+$ and $3/2^+$, respectively, whereas the first three $\Omega_c$ states would be the last $1P$ states with quantum numbers $J^P= 3/2^-$ and $5/2^-$, in that order. Interesting discussions on this quark model picture can be found in Ref.~\cite{OmegacKarlinerRosner} and similar approaches in Refs.~\cite{OmegacMaiani,OmegacXiangLiu,OmegacWeiWang,OmegacQianZhao}.



Other methods have also been employed to study these states, as QCD Sum Rules in Refs.~\cite{azizi,chenhosaka,zhiwang,aliev,sundu,maohosaka,Wang:2017xam} and Lattice QCD in Ref.~\cite{OmegacLattice}.
Pentaquark options have been suggested in Refs.~\cite{yang,huangping,polyakov,an,sheldon,anisovich}. Some works have emphasized the value of decay properties to obtain information on the nature of these states \cite{zezhao,wangzhao,kim} and a discussion on the possible quantum numbers was done in Ref.~\cite{chengchiang}.


On the other hand, some of these states could actually be pentaquark-like molecules, dynamically generated from meson-baryon interactions in coupled channels with charm $C=1$, strangeness $S=-2$ and isospin $I=0$. Predictions in the molecular picture using coupled channels of meson-baryon interactions were done in Refs.~\cite{hofmann,tejero,romanets}.
In this picture the interaction in $S$-wave of baryons with spin-parity $J^P=1/2^+$ or $J^P=3/2^+$ with pseudoscalar mesons leads to meson-baryon systems with $J^P=1/2^-$ and $J^P=3/2^-$, respectively. Channels with vector mesons instead of pseudoscalars can also be included resulting in $J^P=1/2^-,\, 3/2^-$ and $5/2^-$. However, most of the recent works adopting this picture manage to relate two or three of the new $\Omega_c$ states to meson-baryon systems with  $J^P=1/2^-$ and $J^P=3/2^-$, dominated by the pseudoscalar-baryon channels.

In Ref.~\cite{romanets} an $SU(6)_{lsf}\times HQSS$ model (HQSS stands for heavy quark spin symmetry) extending the Weinberg-Tomozawa $\pi N$ interaction was employed to make a systematic study of many possible meson-baryon systems. In Ref.~\cite{OmegacPavao} the renormalization scheme of Ref.~\cite{romanets} was reviewed, performing an update of the results of the $C=1$, $S=-2$ and $I=0$ sector in view of the new experimental data. The updated results indicate that one can relate the $\Omega_c(3000)$ to a state with $J^P=1/2^-$ and the $\Omega_c(3050)$ to another state with $J^P=3/2^-$, with hints that the $\Omega_c(3090)$ or $\Omega_c(3119)$ could also have $J^P=1/2^-$.

In Ref.~\cite{tejero} the molecular picture was developed using $SU(4)$ symmetry to extend the interaction described by vector meson exchange in the local hidden gauge approach. This work was also reviewed under the light of the new experimental data and an updated study was made in Ref.~\cite{OmegacMontana}, where it was shown that the $\Omega_c(3050)$ and $\Omega_c(3090)$ can be both related to meson-baryon resonances with $J^P=1/2^-$, stemming from pseudoscalar-baryon($1/2^+$) interaction.

A similar approach which also describes the meson-baryon interaction through vector meson exchange was recently developed in Ref.~\cite{OmegacDebastiani}, using an extension of the local hidden gauge approach \cite{hidden1,hidden2,hidden4,Yama,hideko} and taking into account the spin-flavor wave functions of the baryons. In the present work we will follow the description of the $\Omega_c$ states of Ref.~\cite{OmegacDebastiani}; extensive discussions on the methods and results can be found there, as well as predictions of higher energy states of meson-baryon nature. 
In this framework, which will be discussed in more detail in the next section, the heavy quarks are treated as spectators, what implies that the dominant terms of the interaction come from light vector exchange, therefore the interaction is consistent with $SU(3)$ chiral Lagrangians and heavy quark spin symmetry is preserved for the dominant terms in the $(1/m_Q)$ counting \cite{isgur,neubert,manohar}. A remarkable agreement of both masses and widths of the $\Omega_c(3050)$ and $\Omega_c(3090)$ was obtained from the pseudoscalar-baryon($1/2^+$) interaction, in accordance with results of Ref.~\cite{OmegacMontana}; also an extra sector of pseudoscalar-baryon($3/2^+$) could be related to the $\Omega_c(3119)$, therefore assigning to it the quantum numbers $J^P=3/2^-$.

Other works on the molecular picture followed \cite{Omegac3188,OmegacHosaka,OmegacGeng}. In Ref.~\cite{Omegac3188} the authors propose that the broad structure found around 3188 MeV \cite{OmegacLHCb,OmegacBelle} could be related with a molecular $\Xi D$ state due to the proximity of its threshold around 3185 MeV. As we shall see in the next section, in our approach \cite{OmegacDebastiani} the molecular state dominated by the $\Xi D$ channel corresponds to the $\Omega_c(3090)$. 


It is clear that any constraint on the quantum numbers of these new states is essential to discriminate between the different theoretical models. 
It is possible $-$ or if we dare to say, likely $-$ that some of the new $\Omega_c$ states are $1P$ and/or $2S$ excitations as discussed in the quark model picture, and some are dynamically generated meson-baryon molecules. However, answering these questions would automatically open new ones, since in both pictures, choosing one assignment over the other implies that there should be other partner states, which could be below or above the energy range searched experimentally; some of them might be detectable only in other decay channels. Understanding the properties of the observed states, like the narrowness, the couplings to one channel or another, the presence or absence of certain predicted states, and so on, require more information, not just on the quantum numbers, but also the measurement of such states in different reactions. Radiative transitions could play an important role on this search, such as the confirmation of the feed-down mechanism $\Omega_c(3066) \to \Xi_c^\prime \, \bar K \to \Xi_c \gamma \bar K$, which might be the cause of the enhancement near threshold on the LHCb data \cite{OmegacLHCb}, even tough additional states have not been completely ruled out as an alternative explanation. The search for isospin-violating decays to $\Omega_c \, \pi^0$ or electric dipole transitions to $\Omega_c \, \gamma$ could also bring valuable new information.
In view of this exciting spectroscopy challenge, where any correct explanation brings with itself an equal or greater number of new questions, we will try to shed light on the discussion 
looking for different reactions where the new $\Omega_c$ states could be spotted, and discuss peculiarities of each case that could help distinguish among the various different interpretations on the recent literature.


In the present work we propose the experimental study of these new states through the decay of $\Omega_b^-$ baryons, as suggested in Ref.~\cite{Belyaev}. The mass and lifetime of the $\Omega_b^-$ were recently measured by the LHCb collaboration \cite{OmegabLHCb}, obtaining results compatible with the previous measurements of the same collaboration \cite{OmegabLHCbprevious} and also with the ones of the CDF collaboration \cite{OmegabCDF}, but not with the results of the D{\O} collaboration \cite{OmegabD0}. We shall adopt the mass value listed by the Particle Data Group \cite{pdg2016}, which is quite close to the most recent measurement of LHCb.

We will discuss the decay $\Omega_b^- \to (\Xi_c^+K^-)\pi^-$, which could be performed by the LHCb collaboration \cite{Belyaev} in the future and could be very useful to distinguish states with different structures and quantum numbers. First we present a brief summary of the main results of Ref.~\cite{OmegacDebastiani} and comment on the formalism employed there to obtain the $\Omega_c(3050)$ and $\Omega_c(3090)$ as meson-baryon molecules (further details can be found in the Appendix section). Next we will discuss the $\Omega_b^- \to (\Xi_c^+K^-)\pi^-$ decay and how the coupled channels approach naturally accounts for the dynamical generation of the $\Omega_c$ states from the hadronization that takes place after the conversion of the $b$ quark into a $c$ quark.
Then we show the results of how these two states would be seen in the $\Omega_b^-$ decay if the molecular picture of Ref.~\cite{OmegacDebastiani} is correct, providing solid predictions that could easily be put to proof in the near future.


\section{The $\boldsymbol{\Omega_c(3050)}$ and $\boldsymbol{\Omega_c(3090)}$ in the molecular picture}

In Ref.~\cite{OmegacDebastiani} a thorough discussion was made about the meson-baryon interaction due to the exchange of vector mesons, and an extension of the local hidden gauge approach \cite{hidden1,hidden2,hidden4,Yama,hideko} was used together with a method that takes into account the information of the spin-flavor wave functions of the baryons (see Appendix and Ref.~\cite{OmegacDebastiani} for details). It was shown that considering the heavy quarks as spectators, the interactions can be obtained using only $SU(3)$ symmetry (without the need of $SU(4)$ in the dominant terms), respecting heavy quark spin symmetry in the leading terms where only light vectors are exchanged \cite{isgur,neubert,manohar}.

The procedure begins with the choice of meson-baryon channels in the $C=1$, $S=-2$ and $I=0$ sector, interacting in $S$-wave, with a determined total spin $J$. In the case of $J=1/2$, we can have the pseudoscalar-baryon interactions, where the baryons have $J=1/2$, and also vector-baryon contributions that result in $J=1/2$. However, in Ref.~\cite{OmegacDebastiani} it was shown that the coupling of vector-baryon channels with the pseudoscalar-baryon ones gives only a small contribution that can be safely neglected. Therefore, the states generated from the vector-baryon interaction decouple from the ones with pseudoscalars and can be treated separately. Since they do not play any role in the generation of the $\Omega_c(3050)$ and $\Omega_c(3090)$ in our approach, 
we will leave them aside in the present work. As presented in Ref.~\cite{OmegacDebastiani}, a third state, the $\Omega_c(3119)$, can also be related with another pseudoscalar-baryon system, where the baryons have $J=3/2$, assigning the quantum numbers $J^P=3/2^-$ to the $\Omega_c(3119)$. However, since in our approach this state does not mix with the channels of $J=1/2$, its decay into $\Xi_c \bar K$ calls for additional mechanisms with the exchange of pseudoscalars, which go beyond the scope of our approach here.

The channels chosen for the $J^P=1/2^-$ pseudoscalar-baryon states and their respective threshold masses are shown in Table \ref{tab1}.
\begin{table}[h!]
\caption{$J^P=1/2^-$ states chosen and threshold mass in MeV.}
\centering
\begin{tabular}{c | c c c c }
\hline\hline
{\bf States} ~& ~$\Xi_c\bar{K}$~ & ~$\Xi^{\prime}_c\bar{K}$~ & ~$\Xi D$~ & ~$\Omega_c \eta$ \\
\hline
{\bf Threshold} ~& $2965$ & $3074$ & $3185$ & $3243$ \\
\hline\hline
\end{tabular}
\label{tab1}
\end{table}

Using an extension of the local hidden gauge Lagrangians \cite{hidden1,hidden2,hidden4,Yama,hideko} and the baryon spin-flavor wave functions with the heavy quark as spectator, the pseudoscalar-baryon interaction of each channel due to the exchange of vector mesons was calculated in Ref.~\cite{OmegacDebastiani}, and then used as the kernel of the Bethe-Salpeter equation in the on-shell approximation where the multiple meson-baryon loops are summed up in order to obtain a unitarized scattering amplitude \cite{nsd,ollerulf}
\begin{equation}
\label{Tmatrix}
T = [1 - V G]^{-1}\, V \, ,
\end{equation}
where $V$ is the matrix describing the interaction between every channel (see Appendix for more details) and $G$ is the meson-baryon loop function 
\begin{eqnarray}\label{propagator} 
 G_l &=& i \int \frac{d^4q}{(2\pi)^4} \frac{M_l}{E_l({\bf q})}\frac{1}{k^0+p^0 - q^0 - E_l({\bf q})+i\epsilon} \frac{1}{{\bf q}^2-m^2_l+i\epsilon}\nonumber\\
 &=& \int_{|{\bf q}|< q_{max}} \frac{d^3q}{(2\pi)^3}\frac{1}{2\omega_l({\bf q})}\frac{M_l}{E_l({\bf q})}\frac{1}{k^0+p^0-\omega_l({\bf q}) - E_l({\bf q})+i\epsilon}\, ,
 \end{eqnarray}
where $k^0+p^0=\sqrt{s}$ and $\omega_l$, $E_l$, are the meson and baryon energies in the $l$-channel, respectively, and $m_l,\,M_l$ the meson and baryon masses.
As discussed in Ref.~\cite{OmegacDebastiani} a common sharp cutoff of 650 MeV was used to regularize the propagators. We will adopt the same cutoff value along the present work.


When solving Eq.~\eqref{Tmatrix} as a function of the center-of-mass energy $\sqrt{s}$, one can go to the complex energy plane and look for poles in the second Riemann sheet, which correspond to the dynamically generated states such that the real part of the pole is the mass of the state and its width is given by twice the imaginary part of the pole.
In Table~\ref{poles} we show the poles found that correspond to the $\Omega_c(3050)$ and $\Omega_c(3090)$ states. In addition, we evaluate the couplings $g_i$ of the states obtained to the different channels 
and 
$g_iG_i$, which for $S$-wave gives the strength of the wave function at the origin \cite{OmegacDebastiani,gamermann,acetifun}.
\begin{table}[h!]
\caption{The coupling constants to various channels for the poles in the $J^P={1/2}^-$ sector, with $q_{max}=650$ MeV, and $g_i\,G^{II}_i$ in MeV.}
\centering
\begin{tabular}{c c c c c}
\hline\hline
${\bf 3054.05+i0.44}$ & $\Xi_c\bar{K}$ & $\Xi^{\prime}_c\bar{K}$ & $\Xi D$ & $\Omega_c \eta$ \\
\hline
$g_i$ & $-0.06+i0.14$ & $\bf 1.94+i0.01$ & $-2.14+i0.26$ & $1.98+i0.01$ \\
$g_i\,G^{II}_i$ & $-1.40-i3.85$ & $\bf -34.41-i0.30$ & $9.33-i1.10$ & $-16.81-i0.11$ \\
\hline\hline
${\bf 3091.28+i5.12}$ & $\Xi_c\bar{K}$ & $\Xi^{\prime}_c\bar{K}$ & $\Xi D$ & $\Omega_c \eta$ \\
\hline
$g_i$ & $0.18-i0.37$ & $0.31+i0.25$ & $\bf 5.83-i0.20$ & $0.38+i0.23$ \\
$g_i\,G^{II}_i$ & $5.05+i10.19$ & $-9.97-i3.67$ & $\bf -29.82+i0.31$ & $-3.59-i2.23$ \\
\hline\hline
\end{tabular}
\label{poles}
\end{table}

\section{The $\boldsymbol{\Omega_b^- \to (\Xi_c \, \bar{K}) \, \pi^-}$, $\boldsymbol{\Omega_b^- \to (\Xi_c^\prime \, \bar{K}) \, \pi^-}$
and $\boldsymbol{\Omega_b^- \to (\Xi \, D) \, \pi^-}$ decays}

\begin{figure}[h!]
  \centering
  \includegraphics[width=0.7\textwidth]{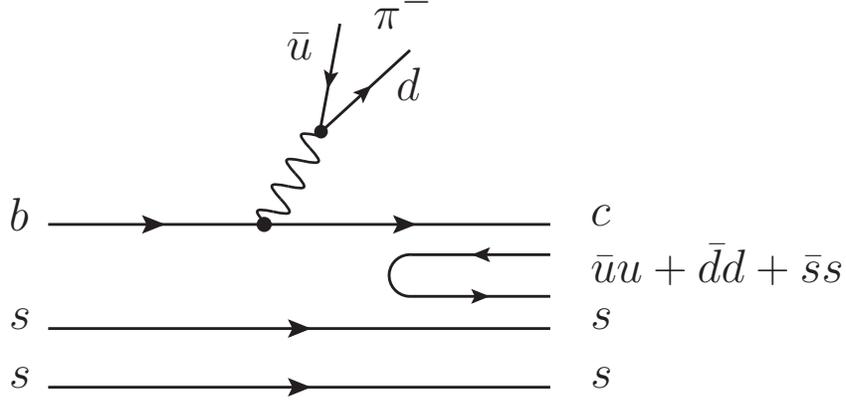}\\
  \caption{$\Omega_b^-$ decay at quark level with emission of a $\pi^-$ and subsequent hadronization.}\label{quarks} 
\end{figure}

As discussed in the previous sections and extensively in Ref.~\cite{OmegacDebastiani}, the $\Omega_c(3050)$ and $\Omega_c(3090)$ are both strong candidates of meson-baryon molecules, dynamically generated from the coupled channels interaction of baryons with $J^P=1/2^+$ and pseudoscalar mesons ($J^P=0^-$) in $S$-wave, therefore in this picture their spin-parity assignment is $J^P=1/2^-$.
Since every quark has intrinsic $J^P=1/2^+$, in order to have $\Xi_c^+K^-$ in $S$-wave in the final state, the hadronization must occur between the $c$ quark and one $s$ quark, as shown in Fig.~\ref{quarks}.

The reason for this approach is that the $b$ quark must turn into a $c$ quark, then both $s$ quarks act as spectators in this decay, so their quantum numbers are fixed a priori as $J^P=1/2^+$, what implies that the $c$ quark must be in $L=1$ to account for the negative parity of the final state, and the hadronization must necessarily involve it so it can be deexcited (see Ref.~\cite{lianghadro} for details). That being said, we include the $\bar{q}{q}$ terms in such a way that the hadronization will occur between the $c$ quark and an $s$ quark.
\begin{align}\label{hadronization}
  css \to c \, (\bar u u + \bar d d + \bar s s)\,  ss \equiv H,\\
  H =  \sum_i c \, \bar q_i q_i \, ss \equiv \sum_i   \displaystyle{\Phi_{4i}} \, q_i ss ,
\end{align}
where in the last step we have identified the $(q\bar{q})$ matrix in Eq.~\eqref{qqbar} with the matrix $\Phi$ of pseudoscalar mesons in Eq.~\eqref{Pmatrix},
\begin{equation}\label{qqbar}
(q \bar q) = \left(
           \begin{array}{cccc}
             u \bar u & u \bar d & u \bar s & u \bar c\\
             d \bar u & d \bar d & d \bar s & d \bar c\\
             s \bar u & s \bar d & s \bar s & s \bar c\\
             c \bar u & c \bar d & c \bar s & c \bar c\\
           \end{array}
            \right) \equiv  \Phi,
\end{equation}
\begin{equation}
\label{Pmatrix}
\Phi =
\left(
\begin{array}{cccc}
\frac{1}{\sqrt{2}} \pi^0 + \frac{1}{\sqrt{3}} \eta  + \frac{1}{\sqrt{6}} \eta^{\prime} & \pi^+ & K^+ & \bar{D}^0 \\
\pi^- &  - \frac{1}{\sqrt{2}} \pi^0 +\frac{1}{\sqrt{3}} \eta + \frac{1}{\sqrt{6}} \eta^{\prime} & K^0 & D^- \\
K^- & \bar{K}^0 & -\frac{1}{\sqrt{3}} \eta +\sqrt{\frac{2}{3}} \eta^{\prime} & D_s^- \\
D^0 & D^+ & D_s^+ & \eta_c\\
\end{array}
\right)\, ,
\end{equation}
where we have included the mixing between $\eta$ and $\eta^{\prime}$ \cite{bramon} for proper matching of $\Phi$ with the $q\bar{q}$ matrix.

Then replacing the $c\bar{q}$ meson terms we get
\begin{equation}\label{H}
  |H\rangle = D^0 \, uss + D^+ dss + \dots
\end{equation}
where we have already neglected the heavy combination of $D_s^+\,sss$ which could only contribute to states with $J^P=3/2^-$ since $sss$ corresponds to the $\Omega^-$. In terms of spin-flavor wave functions, the ground state $\Omega^-$ is flavor symmetric, hence it can only have symmetric spin $\uparrow\uparrow\uparrow$, \textit{i.e.} $J^P=3/2^+$, since the color singlet provides the antisymmetric part of the wave function. Besides, $D_s^+\, \Omega^-$ is far above (threshold at 3641 MeV) the energy range of the channel space considered to generate the meson-baryon molecules.

%

Next, we need to relate the $uss$ and $dss$ with the corresponding baryons, in this case, the $\Xi$ baryons with $J^P=1/2^+$. For consistence with our approach in Ref.~\cite{OmegacDebastiani}, we use the proper spin-flavor wave functions, inspired in Ref.~\cite{Close}, where
\begin{align}
\label{xiwave}
\Xi^0 \equiv \frac{1}{\sqrt{2}}(\phi_{MS}\,\chi_{MS}+\phi_{MA}\,\chi_{MA}),
\end{align}
\vspace{-5pt}
with the Mixed-Symmetric and Mixed-Antisymmetric flavor and spin wave functions
\footnote{Note that the phase convention of $\Xi^0$ is changed in respect to Ref.~\cite{Close}. This is necessary to be consistent with the chiral Lagrangians, as in Table III of Ref.~\cite{miyahara}. See also the footnote of Ref.~\cite{Pavao}.}
:
\begin{align}
\nonumber\phi_{MS} = \frac{1}{\sqrt{6}} [s(us+su) - 2uss]\, ,
\quad
\phi_{MA} = -\frac{1}{\sqrt{2}}[s(us - su)]\, ,
\end{align}
\vspace{-25pt}
\begin{align}
\nonumber\chi_{MS}=\frac{1}{\sqrt{6}}(\uparrow\uparrow\downarrow + \uparrow\downarrow\uparrow - 2 \downarrow\uparrow\uparrow), \,
\quad
\chi_{MA} = \frac{1}{\sqrt{2}}\uparrow\,(\uparrow\downarrow - \downarrow\uparrow\,). \, 
\end{align}

Thus, only the mixed symmetric will contribute and we get the following weight for $\Xi^0$
\begin{align}
\begin{aligned}
  \langle uss | \Xi^0 \rangle &= \frac{1}{\sqrt{6}} \langle uss | sus + ssu -2uss \rangle \\
  &\equiv -\frac{2}{\sqrt{6}}.
\end{aligned}
\end{align}
Equivalently, for the $\Xi^-$ wave function we only replace $u\to d$ and change the sign of $\phi_{MS}$ and $\phi_{MA}$ \cite{OmegacDebastiani}. Again, only the mixed symmetric part will contribute and we get
\begin{align}
\begin{aligned}
  \langle dss | \Xi^- \rangle &= -\frac{1}{\sqrt{6}} \langle dss | sds + ssd -2dss \rangle \\
  &\equiv \frac{2}{\sqrt{6}}.
\end{aligned}
\end{align}

Note that we have dropped the common factor $1/\sqrt{2}$ from Eq.~\eqref{xiwave}, which can be absorbed in the constant factor $V_P$ that we will introduce next.

Then after the hadronization we will have the combination
\begin{equation}\label{combination}
  |H \rangle  = -\frac{2}{\sqrt{6}}  D^0 \, \Xi^0 +\frac{2}{\sqrt{6}}  D^+ \, \Xi^-
  = \frac{2}{\sqrt{3}} | \Xi D, I=0 \rangle ,
\end{equation}
where we have absorbed the global minus sign when we changed to the isospin 0 combination in the order baryon-meson
\footnote{
Recall the isospin doublets:
\begin{equation}\label{DXi}
D = \left(\begin{array}{c} \; D^+\\-D^0 \end{array} \right) , \quad
\Xi = \left(\begin{array}{c} \; \Xi^0\\-\Xi^- \end{array}\right)\, . \nonumber
\end{equation}
} :
\begin{equation}\label{XiDisospin0}
  | \Xi D, I=0 \rangle = -\frac{1}{\sqrt{2}} \Big|\Xi^0D^0 - \Xi^-D^+\Big\rangle\, .
\end{equation}

\begin{figure}[h!]
  \centering
  \includegraphics[width=\textwidth]{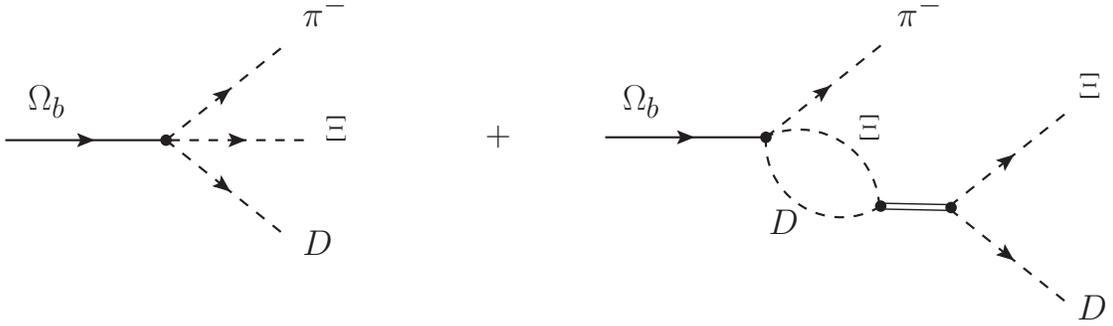}\\
  \caption{$\Omega_b^- \to \pi^- \Xi D$ process. Tree-level (left) plus $\Xi D$ rescattering (right).}\label{diagI}
\end{figure}

Now we can proceed to construct the amplitude of the process $\Omega_b^- \to (\Xi_c \, \bar{K}) \, \pi^-$. It is instructive to first look at the process $\Omega_b^- \to \, (\Xi \, D) \, \pi^- $, as depicted in Fig.~\ref{diagI}. From Eq.~\eqref{combination} we see that after the emission of a pion, the hadronization involving the $css$ quarks generates a $\Xi D$ pair in isospin 0. Thus we can write this process as the sum of a  tree-level contribution and the final state interaction of $\Xi D$ going through the molecular states of Table~\ref{poles}. This information is contained in the diagonal $t$ matrix element $t_{\Xi D\to\Xi D}^{I=0}$, with $t$ the same as in Eq.~\eqref{Tmatrix}, where all the coupled channels dynamics is included. 

Then for the tree-level contribution (left diagram of Fig.~\ref{diagI}) we simply write
\begin{equation}\label{tree}
  t_{\rm tree} = V_P \frac{2}{\sqrt{3}},
\end{equation}
where the weight of $\Xi D$ production comes from Eq.~\eqref{combination} and $V_P$ contains all information related to the $\Omega_b^-$ weak decay, a common unknown factor in all process we will investigate. On the other hand, for the $\Xi  D$ rescattering (right diagram of Fig.~\ref{diagI}) we will have
\begin{equation}\label{loop}
  t_{\rm loop} =  V_P \frac{2}{\sqrt{3}}
G_{\Xi D}[M_{\rm inv}(\Xi D)] \, t_{\Xi D \to \Xi D}^{I=0}[M_{\rm inv}(\Xi D)],
\end{equation}
where $G_{\Xi D}$ is the propagator of the baryon-meson loop, as in Eq.~\eqref{propagator}. Then the amplitude of the process $\Omega_b^- \to \pi^-  \Xi  D $ is given by
\begin{equation}\label{tXiD}
t_{\Omega_b^- \to \pi^- \Xi D} = V_P \frac{2}{\sqrt{3}}
\left( 1 + G_{\Xi D}[M_{\rm inv}(\Xi D)] \, t_{\Xi D \to \Xi D}^{I=0}[M_{\rm inv}(\Xi D)] \right).
\end{equation}
With this amplitude we can write the $\Xi D$ invariant mass distribution
\begin{equation}\label{dGammaXiD}
  \frac{d \Gamma}{d M_{\rm inv}(\Xi D)}
  =\frac{1}{(2\pi)^3}\frac{1}{4M_{\Omega_b^-}^2}\,2M_{\Omega_b^-}\,2M_{\Xi}\, p_{\pi^-} \tilde{p}_{D} \left| t_{\Omega_b^- \to \pi^- \Xi D} \right|^2,
\end{equation}
where $p_{\pi^-}$ is the pion momentum in the $\Omega_b^-$ rest frame
\begin{equation}
p_{\pi^-} = \dfrac{\lambda^{1/2}\left(M^2_{\Omega_b^-}, m^{2}_{\pi},M^2_{\rm inv}(\Xi D) \right)}{2 M_{\Omega_b^-} },
\end{equation}
and $\tilde{p}_D$ is the $D$ momentum in the $\Xi D$ rest frame
\begin{equation}
\tilde{p}_D = \dfrac{\lambda^{1/2}(M^{2}_{\rm inv}(\Xi D), m^2_D,m^2_{\Xi})}{2 M_{\rm inv}(\Xi D)}.
\end{equation}

\begin{figure}[h!]
  \centering
  \includegraphics[width=0.5\textwidth]{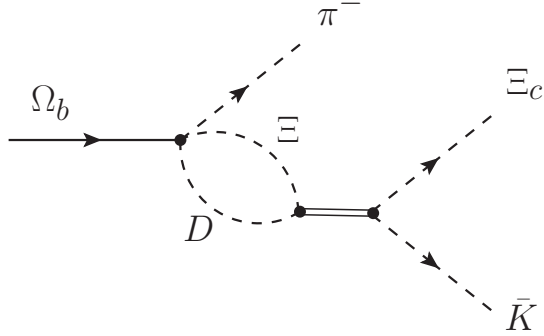}\\
  \caption{$\Omega_b^- \to \pi^- \Xi_c \bar{K}$ process through $\Xi D$ rescattering.}\label{diagII}
\end{figure}

For the $\Omega_b^- \to (\Xi_c \, \bar{K}) \, \pi^-$ process there is no tree-level contribution, since the hadronization only produces a $\Xi D$ pair, as in Eq.~\eqref{combination}, then the only contribution comes from the diagram in Fig.~\ref{diagII}.

Due to our coupled channels approach, the transition $\Xi  D  \to  \Xi_c  \bar K$ is already contained in the $t$ matrix and the production of $\Xi_c  \bar K$ (also in isospin 0) appears naturally. The corresponding amplitude will be
\begin{equation}\label{tKbarXic}
t_{\Omega_b^- \to \pi^-  \Xi_c \bar K} = V_P \frac{2}{\sqrt{3}}
G_{\Xi D}[M_{\rm inv}(\Xi_c \bar K)] \, t_{\Xi D\to\Xi_c \bar K}[M_{\rm inv}(\Xi_c \bar K)],
\end{equation}
where $t_{\Xi D\to\Xi_c \bar K}$ is the transition amplitude of $\Xi  D \to \Xi_c  \bar K$, from the same $t$ matrix. Then the $\Xi_c \bar K$ invariant mass distribution is also analogous,
\begin{equation}\label{dGammaXicKbar}
  \frac{d \Gamma}{d M_{\rm inv}(\Xi_c \bar K)}
  =\frac{1}{(2\pi)^3}\frac{1}{4M_{\Omega_b^-}^2}\,2M_{\Omega_b^-}\,2M_{\Xi_c}\, p_{\pi^-} \tilde{p}_{\bar K} \left| t_{\Omega_b^- \to \pi^- \Xi_c \bar K} \right|^2,
\end{equation}
where
\begin{equation}
p_{\pi^-} = \dfrac{\lambda^{1/2}\left(M^2_{\Omega_b^-}, m^{2}_{\pi},M^2_{\rm inv}(\Xi_c \bar K) \right)}{2 M_{\Omega_b^-} },
\end{equation}
and
\begin{equation}\label{ptildeK}
\tilde{p}_{\bar K} = \dfrac{\lambda^{1/2}(M^{2}_{\rm inv}(\Xi_c \bar K), m^2_{K},M^2_{\Xi_c})}{2 M_{\rm inv}(\Xi_c \bar K)}.
\end{equation}

Analogously, we can also calculate the invariant mass distribution for the final state $\Xi_c^\prime \bar K$, replacing $\Xi_c$ by $\Xi_c^\prime$ in the previous equations and taking the matrix element of the $t$ matrix corresponding to the transition $\Xi D \to \Xi_c^\prime \bar K$.

\begin{figure}[h!]
  \centering
  \includegraphics[width=0.5\textwidth]{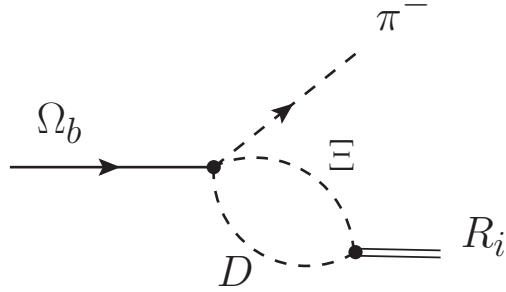}\\
  \caption{Resonance coalescence in the $\Omega_b^- \to \pi^- R_i$ process through $\Xi D$ rescattering, where $R_i$ is the $\Omega_c(3050)$ or $\Omega_c(3090)$.}\label{diagIII}
\end{figure}

It is also interesting to look at the case of coalescence, where the $\Xi  D$ pair merge into the resonance regardless of the final decay channel, as depicted in Fig.~\ref{diagIII}.

The value of the amplitude in the process $\Omega_b^- \to \pi^- R_i$, where $R_i$ is one of the molecular states of Table~\ref{poles}, is proportional to the coupling of that resonance to the $\Xi D$ channel,
\begin{equation}\label{tRi}
t_{\Omega_b^- \to \pi^- R_i} = V_P \frac{2}{\sqrt{3}}
G_{\Xi D}(M_{R_i}) g_{R_i,\Xi D} \, ,
\end{equation}
where the propagator is calculated at the resonance mass $M_{R_i}$. With this quantity we can calculate the equivalent of the integrated mass distribution around the $R_i$ resonance, which does not depend on its decay mode,
\begin{equation}\label{GammaRi}
  \Gamma_{\Omega_b^- \to \pi^- R_i}
  =\frac{1}{8\pi}\frac{1}{M_{\Omega_b^-}^2}\,2M_{\Omega_b^-}\,2M_{R_i}\, p^\prime_{\pi^-} \left| t_{\Omega_b^- \to \pi^- \Xi_c \bar K}(M_{R_i}) \right|^2,
\end{equation}
where
\begin{equation}
p^\prime_{\pi^-} = \dfrac{\lambda^{1/2}\left(M^2_{\Omega_b^-}, m^{2}_{\pi},M^2_{R_i} \right)}{2 M_{\Omega_b^-} }.
\end{equation}

\section{Results}

\begin{figure}[h!]
  \centering
  \includegraphics[width=0.8\textwidth]{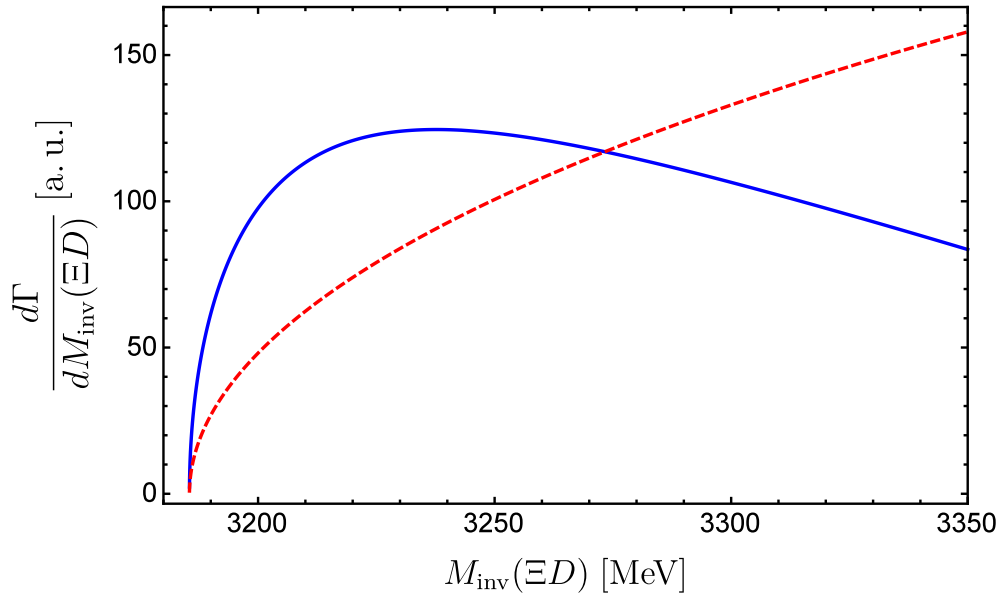}\\
  \caption{$\Xi D$ invariant mass distribution from Eq.~\eqref{dGammaXiD}. Solid line: using the complete amplitude of Eq.~\eqref{tXiD}. Dashed line: removing the $G_{\Xi D} \,t_{\Xi D \to \Xi D}^{I=0}$ term (only tree-level contribution) and normalizing such that both curves have the same area.}\label{XiD}
\end{figure}

It is interesting to see how these processes show the importance of the coupled channels. 
The decay of the $\Omega_c$ states into $\Xi D$ is kinematically forbidden below the corresponding threshold at 3185 MeV, then we cannot see the corresponding peaks in the $\Xi D$ invariant mass distribution, but we can see their indirect effect, both from the meson-baryon loop in Eq.~\eqref{tXiD} and from the amplitude $t_{\Xi D \to \Xi D}^{I=0}$. The corresponding invariant mass distribution, Eq.~\eqref{dGammaXiD}, is shown in Fig.~\ref{XiD} by the solid line. To compare with the case where only tree-level contributes, we remove $t_{\rm loop}$ and keep only $t_{\rm tree}$, normalizing the curve such that it has the same area as the solid curve in the energy range shown, which is plotted as the dashed line in Fig.~\ref{XiD}.

The $\Xi_c \bar K$ threshold is at 2965 MeV, then we can see clearly, of course, the peaks of the $\Omega_c$ states in the $\Xi_c \bar K$ invariant mass distribution.
According to Eq.~\eqref{combination}, we expect only $\Xi D$ production from the hadronization that occurs right after the $\Omega_b^-$ decay, which means we have no tree-level contribution for $\Xi_c \bar K$ production. However, the transition to
$\Xi_c \bar{K}$ through off-shell $\Xi D$ loops arises naturally from the coupled channels approach.
In fact, both the $\Omega_c(3050)$ and $\Omega_c(3090)$ couple strongly to $\Xi D$ (see Table \ref{poles}), and their formation from the $\Xi D$ state formed in the first step of the $\Omega_b^-$ decay with subsequent transition to $\Xi_c \bar{K}$ (going through the $\Xi D$ virtual state) is not only possible, but expected.

\begin{figure}[h!]
  \centering
  \includegraphics[width=0.8\textwidth]{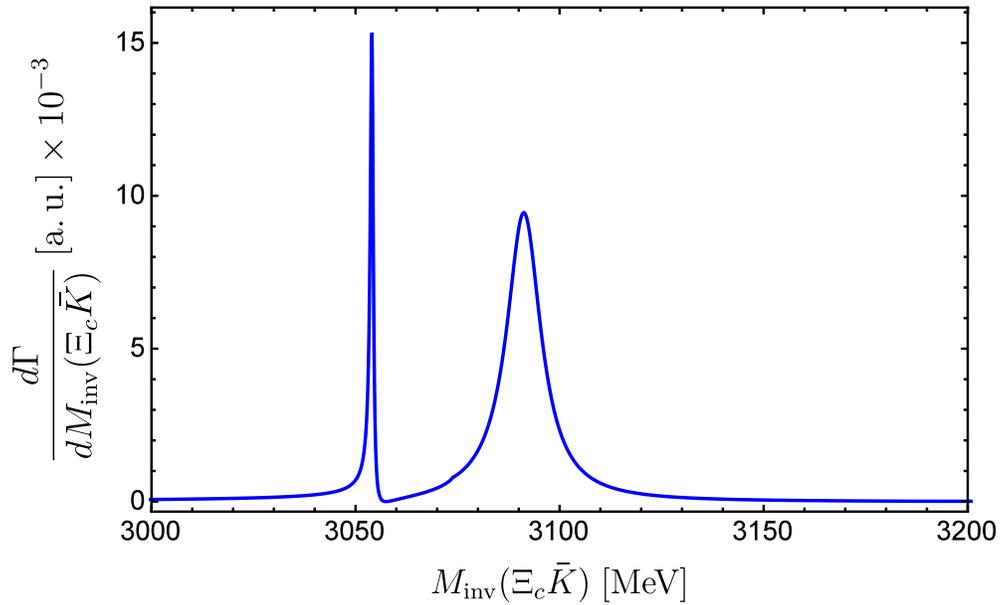}\\
  \caption{$\Xi_c \bar{K}$ invariant mass distribution from Eq.~\eqref{dGammaXicKbar}.}\label{XicKbar}
\end{figure}

In Fig.~\ref{XicKbar} we show the $\Xi_c \bar{K}$ invariant mass distribution. The only unknown quantity is the global factor $V_P$, common to all amplitudes we investigate here. The ratio between the intensity of each peak does not depend on $V_P$, so all ratios are predictions that could be confronted with future experiments. We can see that the intensity of the $\Omega_c(3050)$ peak is about 65\% higher than the $\Omega_c(3090)$ peak.
Note that we are using the same normalization for all the reactions, hence the ratio of the strength at the peaks in Fig.~\ref{XicKbar} to the strength of the $\Xi D$ mass distribution (solid line) in Fig.~\ref{XiD} is also a prediction.
In arbitrary units, the $\Xi D$ distribution has a maximum of about 125 around 3240 MeV, whereas in the $\Xi_c \bar{K}$ distribution the $\Omega_c(3050)$ and $\Omega_c(3090)$ peaks have an intensity of about $15.50 \times 10^3$ and $9.45 \times 10^3$, respectively, therefore we predict that the $\Xi_c \bar{K}$ distribution in the vicinity of the resonances peaks is roughly two orders of magnitude higher than the $\Xi D$ distribution.

Cusp effects in the $\Xi_c \bar{K}$ distribution also appear at the $\Xi_c^\prime \bar{K}$ and $\Xi D$ thresholds at 3074 MeV and 3185 MeV, respectively, but their intensity is very small compared to the peaks of the resonances and cannot be seen clearly in Fig.~\ref{XicKbar}. 

We also notice that apparently no significant interference pattern is seen between both states, even tough they have the same quantum numbers, a feature that also agrees with the fit performed by the LHCb collaboration \cite{OmegacKarlinerRosner,OmegacLHCb}.

\begin{figure}[h!]
  \centering
  \includegraphics[width=0.8\textwidth]{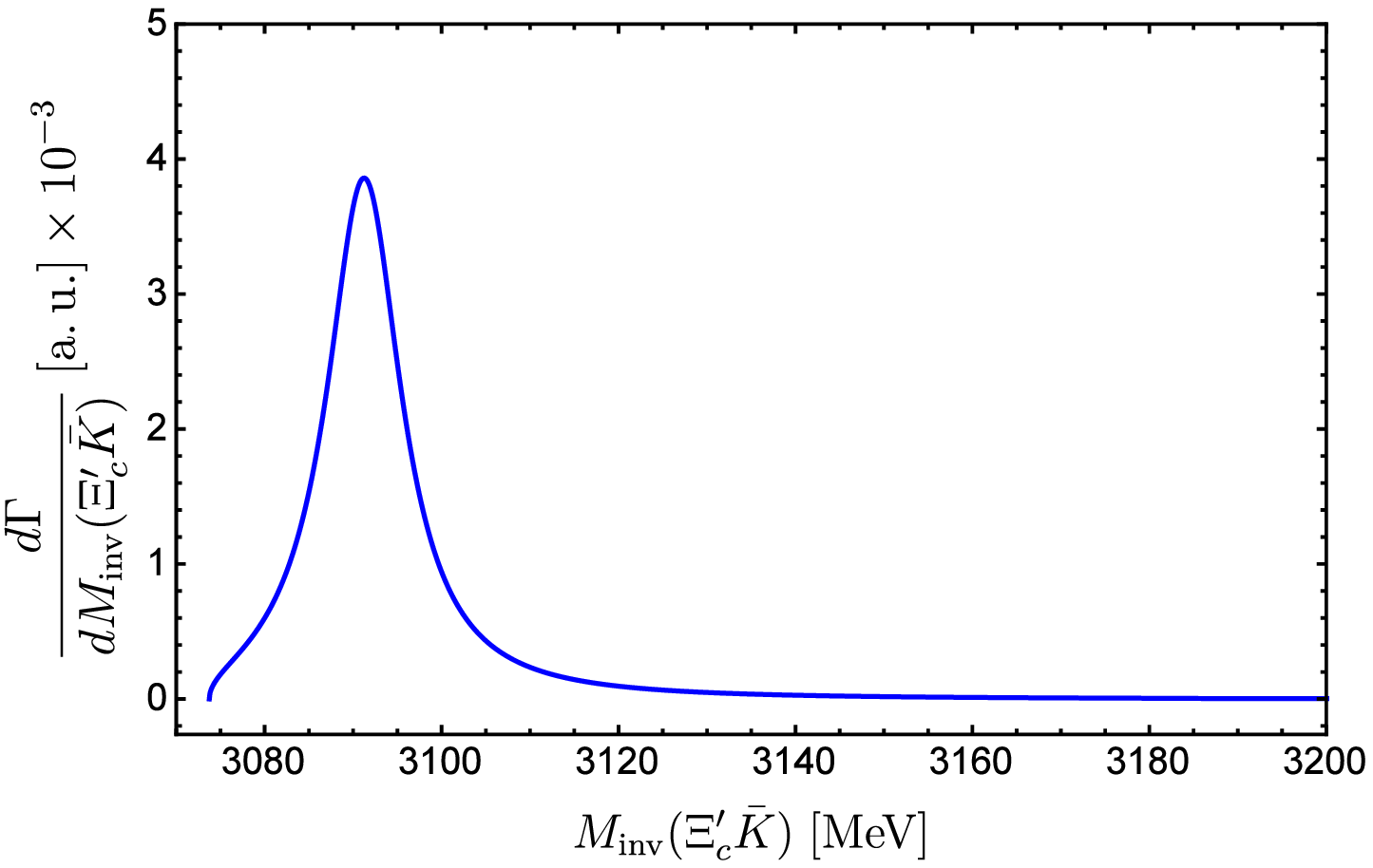}\\
  \caption{$\Xi_c^\prime \bar{K}$ invariant mass distribution from Eq.~\eqref{dGammaXicKbar}, replacing $\Xi_c$ by $\Xi_c^\prime$ and taking $t_{\Xi D \to \Xi_c^\prime \bar K}$.}\label{XicPrimeKbar}
\end{figure}

As for the $\Xi_c^\prime  \bar K$ invariant mass distribution, only the $\Omega_c(3090)$ can be seen, as shown in Fig.~\ref{XicPrimeKbar}, since this channel is also open for the decay of this state, whereas the $\Omega_c(3050)$ is below the threshold of $\Xi_c^\prime \bar K$.

Again we can compare the intensity of the peaks. In the $\Xi_c^\prime \bar{K}$ distribution the $\Omega_c(3090)$ peak has an intensity of $3.86\times10^3$, which is about 40\% of the intensity it has in the $\Xi_c \bar{K}$. 

Finally, it is interesting to compare the results of $\Gamma_{\Omega_b^- \to \pi^- R_i}$, given by Eq.~\eqref{GammaRi}, with the integrated invariant mass distribution of Eq.~\eqref{dGammaXicKbar} around the peak of each resonance,
\begin{equation}\label{intGammaRi}
  \int_{R_i}
  \frac{d \Gamma}{d M_{\rm inv}(\Xi_c \bar K)}\,
  d M_{\rm inv}(\Xi_c \bar K).
\end{equation}

In Table \ref{tabCoales} we show the results of Eqs.~\eqref{GammaRi} and \eqref{intGammaRi} for the $\Omega_c(3050)$ and $\Omega_c(3090)$, and for the latter we also show the integrated $\Xi_c^\prime \bar K$ distribution.
\begin{table}[h!]
\caption{Comparison between integrated invariant mass distributions around each resonance and the corresponding coalescence.\label{tabCoales}}
\centering
\begin{tabular}{c | c | c | c| c}
\hline\hline
 State & \multicolumn{2}{c|}{$\Omega_c(3050)$} & \multicolumn{2}{c}{$\Omega_c(3090)$} \\
\hline
 Pole & \multicolumn{2}{c|}{${\bf 3054.05+i0.44}$} & \multicolumn{2}{c}{${\bf 3091.28+i5.12}$} \\
\hline
Coalescence Eq.~\eqref{GammaRi} [a. u.] & \multicolumn{2}{c|}{$21 289$} & \multicolumn{2}{c}{$215 237$} \\
\hline
Channel & \multicolumn{2}{c|}{$\Xi_c \bar K$} & $\Xi_c \bar K$ & $\Xi_c^\prime \bar K$  \\
\hline
Interval [MeV] & \multicolumn{2}{c|}{$[3049 , 3057]$}  & $[3057 , 3120]$ & $[3074 , 3120]$ \\
\hline
Integral Eq.~\eqref{intGammaRi} [a. u.] & \multicolumn{2}{c|}{$21 344$}  & $133 482$ & $51 074$ \\
\hline\hline
\end{tabular}
\end{table}

From these results we can draw some interesting conclusions. Let us look first at the $\Omega_c(3050)$. In Table \ref{poles} we can see that this state is dominated by the $\Xi_c^\prime \bar K$ channel, and also has sizable contributions from the higher channels $\Xi D$ and $\Omega_c \eta$. However, the pole is at 3054 MeV, which is 20 MeV below the $\Xi_c^\prime \bar K$ threshold, so the only open channel is $\Xi_c \bar K$, to which the resonance couples very weakly and the phase space available is only about 90 MeV. This feature explains two points: 1) The narrowness of the state, whose upper limit of the width reported by the LHCb collaboration is $0.8 \pm 0.2 \pm 0.1$ MeV \cite{OmegacLHCb}, in excellent agreement with our result: $\Gamma = 2\times {\rm Im}(R_i) = 0.88$ MeV; 2) The good agreement of $\Gamma_{\Omega_b^- \to \pi^- R_i}$, $21 289$, given by Eq.~\eqref{GammaRi}, with the integrated invariant mass distribution around the resonance peak, $21 344$, given by Eq.~\eqref{intGammaRi} (the small difference is irrelevant and comes essentially from the choice of the interval of integration). This happens because the state is very narrow and the only channel open for decay is the $\Xi_c \bar K$, so the value we obtain from the coalescence (which is independent of the decay channel), matches the value obtained from the integration over the only channel available ($\Xi_c \bar K$) as it should be.

On the other hand, the $\Omega_c(3090)$ is dominated by the $\Xi D$ channel, with some contribution from the other channels. Even tough this state couples very strongly to $\Xi D$, it is almost 100 MeV below the respective threshold. It can only decay to $\Xi_c \bar K$ and $\Xi_c^\prime \bar K$. In both cases the coupling is small, and in the latter channel the phase space available is less than 20 MeV (see Table \ref{poles}). This again, explains two main features: 1) The narrowness, with a width not so small as in the case of the $\Omega_c(3050)$, but still very narrow, since the decay into $\Xi_c \bar K$ is reasonable, with more than 120 MeV of phase space available, although the coupling to that channel is weak. The LHCb reports a width of $8.7 \pm 1.0 \pm 0.8$ MeV \cite{OmegacLHCb}, in fair agreement with our result of $10.24$ MeV; 2) The fact that the integrated invariant mass distribution around the peak, $133 482$ in the $\Xi_c \bar K$ distribution, is about $2/3$ of the total given by the coalescence, $215 237$, which is also expected, since in Eq.~\eqref{intGammaRi} we are integrating only in the $\Xi_c \bar K$ channel, whereas this state can also decay into $\Xi_c^\prime \bar K$. The sum of both integrals, in $\Xi_c \bar K$ and $\Xi_c^\prime \bar K$, is $184 556$, close to the total given by the coalescence, but still below. This is also expected since Eq.~\eqref{GammaRi} is actually an approximation that is valid in the limit of zero width, which works pretty well for the $\Omega_c(3050)$ but is not so good for the $\Omega_c(3090)$, with already 10 MeV of width. This can also be verified calculating the contribution of each channel to the total width, in terms of the coupling of the resonance to that channel. For $\Xi_c \bar K$ \vspace{-1pt}
\begin{equation}\label{Gam3090}
  \Gamma_{R_i, \, \Xi_c \bar K} = \frac{1}{8\pi}\frac{1}{M_{R_i}^2}
  \,2M_{R_i}\,2M_{\Xi_c}\,
  |g_{R_i, \Xi_c \bar K}|^2\, \tilde{p_2} \, ,
\end{equation}
where $M_{R_i}$ is the mass of the resonance and $\tilde{p_2}$ is given by
\begin{equation}
\tilde{p_2} = \dfrac{\lambda^{1/2}(M_{R_i}^2, m_K^2, M^2_{\Xi_c})}{2M_{R_i}}.
\end{equation}

Using this equation we get $\Gamma_{\Xi_c \bar K}=0.83$ MeV for the $\Omega_c(3050)$, in accordance with the value obtained from the pole position. For the $\Omega_c(3090)$ we get $\Gamma_{\Xi_c \bar K}=7.24$ MeV for the $\Xi_c \bar K$, and replacing $\Xi_c$ by $\Xi_c^\prime$ and the corresponding coupling, we get $\Gamma_{\Xi_c^\prime \bar K}=2.56$ MeV, which add up to $9.80$ MeV, a bit below $10.24$ MeV that we get from the pole position. 

As a prediction we can also state, based on the coalescence results, that the ratio of the $\Omega_c(3050)$ over the $\Omega_c(3090)$ production is about 10\% in the $\Omega_b^-$ decay:
\begin{equation}
    \frac{\Gamma_{\Omega_b^- \to \pi^- \Omega_c(3050)}}{\Gamma_{\Omega_b^- \to \pi^- \Omega_c(3090)}} \approx 10\%.
\end{equation}

\section{Conclusions}

We have studied the weak decay $\Omega_b^- \to (\Xi_c^+ \, K^-) \, \pi^-$, in view of the narrow $\Omega_c$ states recently measured by the LHCb collaboration and later confirmed by the Belle collaboration. Based on the previous work where the $\Omega_c(3050)$ and $\Omega_c(3090)$ are described as meson-baryon molecular states, using an extension of the local hidden gauge approach in coupled channels, with results in remarkable agreement with experiment, we have investigated the $\Xi D$, $\Xi_c \bar K$ and $\Xi_c^\prime \bar K$ invariant mass distributions, and discussed the role of coupled channels in the process. Predictions that could be confronted with future experiments are presented, providing useful information that could help to determine the quantum numbers and nature of these states. Since $\Omega_b^-$ baryons have already been observed in several experiments, two of them performed by the LHCb collaboration, the present work should encourage such study in the near future, which would certainly bring novel key information for the understanding of these new states.


\begin{acknowledgments}
V. R. Debastiani acknowledges the Programa Santiago Grisolia of Generalitat Valenciana (Exp. GRISOLIA/2015/005) and the discussions with I. Belyaev and S. Neubert. J. M. Dias thanks the Funda\c c\~ao de Amparo \`a Pesquisa do Estado de S\~ao Paulo (FAPESP) for support by FAPESP grant 2016/22561-2. This work is partly supported by the National Natural Science Foundation of China (NSFC) under Grant Nos. 11565007 and 11747307. This work is also partly supported by the Spanish Ministerio de Economia y Competitividad and European FEDER funds under the contract number FIS2014-57026-REDT, FIS2014-51948-C2-1-P, and FIS2014-51948-C2-2-P, and the Generalitat Valenciana in the program Prometeo II-2014/068.
\end{acknowledgments}

\section*{Appendix}

The ingredients needed to obtain the molecular states presented here are the vector(V)-pseudoscalar(P)-pseudoscalar(P) Lagrangian from the local hidden gauge approach,
\begin{eqnarray}
\label{vppLag}
\mathcal{L}_{VPP} = -i g\, \langle \,\, [\,\Phi,\,\partial_{\mu} \Phi\,]\, V^{\mu}\, \rangle \, ,
\end{eqnarray}
for the VPP vertex, where $g=m_V/2f_\pi$; $f_\pi=93$ MeV is the pion decay constant, and $m_V$ is the mass of the light vector mesons. $\Phi$ is the pseudoscalar mesons matrix from Eq.~\eqref{Pmatrix} and $V^\mu$ is the vector mesons matrix,
\begin{equation}
\label{Vmatrix}
V^\mu =
\left(
\begin{array}{cccc}
\frac{1}{\sqrt{2}} \rho^0 + \frac{1}{\sqrt{2}} \omega & \rho^+ & K^{* +} & \bar{D}^{* 0} \\
\rho^- & -\frac{1}{\sqrt{2}} \rho^0 + \frac{1}{\sqrt{2}} \omega & K^{* 0} & \bar{D}^{* -} \\
K^{* -} & \bar{K}^{* 0} & \phi & D_s^{* -} \\
D^{* 0} & D^{* +} & D_s^{* +} & J/\psi\\
\end{array}
\right)\, .
\end{equation}
For the VBB vertex, instead of extending from $SU(3)$ to $SU(4)$ the vector(V)-baryon(B)-baryon(B) Lagrangian from the local hidden gauge approach, we construct spin-flavor wave functions for the baryons, considering the heavy quark as spectator and symmetrizing only in the light quarks. Then we write the vector mesons in terms of their quark content and apply them in the wave functions as number operators times the coupling constant from vector meson exchange (the same of Eq.~\eqref{vppLag}).
As discussed in Ref.~\cite{OmegacDebastiani}, this method yields the same results as the VBB Lagrangian of the local hidden gauge in $SU(3)$, being consistent at the same time with this formalism and with the hypothesis of the heavy quark as spectator, ensuring heavy quark spin symmetry in the dominant terms where only light vectors are exchanged.
%

With these ingredients and the baryon spin-flavor wave functions, one can obtain the coefficients $D_{ij}$ of the $V$ matrix that we plug in the Bethe-Salpeter equation in the form of Eq.~\eqref{Tmatrix}, given by
\begin{equation}
    \label{vijRelat}
V_{ij}= D_{ij}\frac{2\sqrt{s} -M_{B_i} -M_{B_j}}{4f_\pi^2}  \sqrt{\frac{M_{B_i} + E_{B_i}}{2M_{B_i}}} \sqrt{\frac{M_{B_j} + E_{B_j}}{2M_{B_j}}}\, ,
\end{equation}
where $M_{{B_i},{B_j}}$ and $E_{{B_i},{B_j}}$ stand for the mass and the center-of-mass energy of the baryons, respectively, and the matrix $D_{ij}$ is given in Table \ref{tabDij}.
\begin{table}[h!]
\caption{$D_{ij}$ coefficients of Eq.~\eqref{vijRelat} for the $J^P={1/2}^-$ meson-baryon states in $S$-wave.\label{tabDij}}
\centering
\begin{tabular}{c || c c c c}
\hline\hline
 $J=1/2$~ & ~~$\Xi_c\bar{K}$~ & ~$\Xi^{\prime}_c\bar{K}$~ & ~$\Xi D$~ & ~$\Omega_c \eta$ \\
\hline\hline
$\Xi_c\bar{K}$ & $-1$ & $0$ & $-\frac{1}{\sqrt{2}}\lambda$ \\
$\Xi^{\prime}_c\bar{K}$ &  & $-1$ & $\frac{1}{\sqrt{6}}\lambda$ & $-\frac{4}{\sqrt{3}}$ \\
$\Xi D$ & & & $-2$ & $\frac{\sqrt{2}}{3}\lambda$ \\
$\Omega_c \eta$ & & &  & $0$ \\
\hline\hline
\end{tabular}
\end{table}

The details of the calculation can be found in Ref.~\cite{OmegacDebastiani}. In Table \ref{tabDij} we have the parameter $\lambda$ in some non diagonal matrix elements, which involve transitions from one meson without charm to one with charm, like $\bar{K} \to D$. In this case we have to exchange a heavy quark, and the propagator of the exchanged vector goes like
\begin{eqnarray}
\frac{1}{(q^0)^2 - |{\bf q\,}|^2-m_{D^*_s}^2} \approx \frac{1}{(m_D - m_K)^2-m_{D^*_s}^2}\, ,
\end{eqnarray}
and the ratio to the propagator of the light vectors is
\begin{eqnarray}
\lambda \equiv \frac{-m^2_{V}}{(m_D - m_K)^2 - m^2_{D^*_s}}\approx 0.25 \, .
\end{eqnarray}
Then we take $\lambda = 1/4$ in all these matrix elements, as it was done in Ref.~\cite{mizutani}.

The diagonal matrix elements of Table \ref{tabDij} coincide with those of Ref.~\cite{OmegacMontana},
but not all the non diagonal. $\textrm{SU(4)}$ symmetry is used in Ref.~\cite{OmegacMontana},
but only $\textrm{SU(3)}$ is effectively used in the diagonal terms. The same happens in our approach, with the difference that the heavy baryon wave functions we have constructed are not eigenstates of $\textrm{SU(4)}$, since we treat the heavy quark as an spectator and use $SU(3)$ for the light quarks. This induces a spin-flavor dependence different from the one of pure $\textrm{SU(4)}$ symmetry.



\begin{thebibliography}{99}

%
\bibitem{OmegacLHCb}
  R.~Aaij {\it et al.} [LHCb Collaboration],
  ``Observation of five new narrow $\Omega_c^0$ states decaying to $\Xi_c^+ K^-$,''
  Phys.\ Rev.\ Lett.\  {\bf 118}, 182001 (2017).

%
\bibitem{OmegacBelle}
  J.~Yelton {\it et al.} [Belle Collaboration],
  ``Observation of Excited $\Omega_c$ Charmed Baryons in $e^+e^-$ Collisions,''
  arXiv:1711.07927 [hep-ex].



%
\bibitem{ebert}
  D.~Ebert, R.~N.~Faustov and V.~O.~Galkin,
  ``Masses of excited heavy baryons in the relativistic quark model,''
  Phys.\ Lett.\ B {\bf 659}, 612 (2008).

%
\bibitem{roberts}
  W.~Roberts and M.~Pervin,
  ``Heavy baryons in a quark model,''
  Int.\ J.\ Mod.\ Phys.\ A {\bf 23}, 2817 (2008).

%
\bibitem{garcilazo}
  H.~Garcilazo, J.~ and A.~Valcarce,
  ``Faddeev study of heavy baryon spectroscopy,''
  J.\ Phys.\ G {\bf 34}, 961 (2007).

%
\bibitem{migura}
  S.~Migura, D.~Merten, B.~Metsch and H.~R.~Petry,
  ``Charmed baryons in a relativistic quark model,''
  Eur.\ Phys.\ J.\ A {\bf 28}, 41 (2006).

%
\bibitem{ebert2}
  D.~Ebert, R.~N.~Faustov and V.~O.~Galkin,
  ``Spectroscopy and Regge trajectories of heavy baryons in the relativistic quark-diquark picture,''
  Phys.\ Rev.\ D {\bf 84}, 014025 (2011).

%
\bibitem{valcarce}
  A.~Valcarce, H.~Garcilazo and J.~Vijande,
  ``Towards an understanding of heavy baryon spectroscopy,''
  Eur.\ Phys.\ J.\ A {\bf 37}, 217 (2008).

%
\bibitem{shah}
  Z.~Shah, K.~Thakkar, A.~K.~Rai and P.~C.~Vinodkumar,
  ``Mass spectra and Regge trajectories of $\Lambda_{c}^{+}$, $\Sigma_{c}^{0}$, $\Xi_{c}^{0}$ and $\Omega_{c}^{0}$ baryons,''
  Chin.\ Phys.\ C {\bf 40}, no. 12, 123102 (2016).

%
\bibitem{vijande}
  J.~Vijande, A.~Valcarce, T.~F.~Carames and H.~Garcilazo,
  ``Heavy hadron spectroscopy: a quark model perspective,''
  Int.\ J.\ Mod.\ Phys.\ E {\bf 22}, 1330011 (2013).

%
\bibitem{yoshida}
  T.~Yoshida, E.~Hiyama, A.~Hosaka, M.~Oka and K.~Sadato,
  ``Spectrum of heavy baryons in the quark model,''
  Phys.\ Rev.\ D {\bf 92}, no. 11, 114029 (2015).

%
\bibitem{cheng}
  H.~X.~Chen, W.~Chen, Q.~Mao, A.~Hosaka, X.~Liu and S.~L.~Zhu,
  ``P-wave charmed baryons from QCD sum rules,''
  Phys.\ Rev.\ D {\bf 91}, no. 5, 054034 (2015).

%
\bibitem{cheng2}
  H.~X.~Chen, Q.~Mao, A.~Hosaka, X.~Liu and S.~L.~Zhu,
  ``D-wave charmed and bottomed baryons from QCD sum rules,''
  Phys.\ Rev.\ D {\bf 94}, no. 11, 114016 (2016).

%
\bibitem{chiladze}
  G.~Chiladze and A.~F.~Falk,
  ``Phenomenology of new baryons with charm and strangeness,''
  Phys.\ Rev.\ D {\bf 56}, R6738 (1997).

\bibitem{chiralquark}
  A.~Manohar and H.~Georgi,
  ``Chiral Quarks and the Nonrelativistic Quark Model,''
  Nucl.\ Phys.\ B {\bf 234}, 189 (1984).



%
\bibitem{OmegacKarlinerRosner}
  M.~Karliner and J.~L.~Rosner,
  ``Very narrow excited $\Omega_c$ baryons,'',
  Phys.\ Rev.\ D {\bf 95}, no. 11, 114012 (2017).





%
\bibitem{OmegacXiangLiu}
  B.~Chen and X.~Liu,
  ``New $\Omega_c^0$ baryons discovered by LHCb as the members of $1P$ and $2S$ states,''
  Phys.\ Rev.\ D {\bf 96}, no. 9, 094015 (2017).


%
\bibitem{OmegacWeiWang}
  W.~Wang and R.~L.~Zhu,
  ``Interpretation of the newly observed $\Omega_c^0$ resonances,''
  Phys.\ Rev.\ D {\bf 96}, no. 1, 014024 (2017).


%
\bibitem{OmegacQianZhao}
  K.~L.~Wang, L.~Y.~Xiao, X.~H.~Zhong and Q.~Zhao,
  ``Understanding the newly observed $\Omega_c$ states through their decays,''
  Phys.\ Rev.\ D {\bf 95}, no. 11, 116010 (2017).


%
\bibitem{OmegacMaiani}
  A.~Ali, L.~Maiani, A.~V.~Borisov, I.~Ahmed, M.~Jamil Aslam, A.~Y.~Parkhomenko, A.~D.~Polosa and A.~Rehman,
  ``A new look at the Y tetraquarks and $\Omega _c$ baryons in the diquark model,''
  Eur.\ Phys.\ J.\ C {\bf 78}, no. 1, 29 (2018).




%
\bibitem{azizi}
  S.~S.~Agaev, K.~Azizi and H.~Sundu,
  ``On the nature of the newly discovered $\Omega$ states,''
  EPL {\bf 118}, no. 6, 61001 (2017).

%
\bibitem{chenhosaka}
  H.~X.~Chen, Q.~Mao, W.~Chen, A.~Hosaka, X.~Liu and S.~L.~Zhu,
  ``Decay properties of $P$-wave charmed baryons from light-cone QCD sum rules,''
  Phys.\ Rev.\ D {\bf 95}, no. 9, 094008 (2017).

%
\bibitem{zhiwang}
  Z.~G.~Wang,
  ``Analysis of $\Omega _c(3000)$ , $\Omega _c(3050)$ , $\Omega _c(3066)$ , $\Omega_c(3090)$ and $\Omega _c(3119)$ with QCD sum rules,''
  Eur.\ Phys.\ J.\ C {\bf 77}, no. 5, 325 (2017).

%
\bibitem{aliev}
  T.~M.~Aliev, S.~Bilmis and M.~Savci,
  ``Are the new excited $\Omega_c$ baryons negative parity states?,''
  arXiv:1704.03439 [hep-ph].

%
\bibitem{sundu}
  S.~S.~Agaev, K.~Azizi and H.~Sundu,
  ``Interpretation of the new $\Omega_c^{0}$ states via their mass and width,''
  Eur.\ Phys.\ J.\ C {\bf 77}, no. 6, 395 (2017).

%
\bibitem{maohosaka}
  Q.~Mao, H.~X.~Chen, A.~Hosaka, X.~Liu and S.~L.~Zhu,
  ``$D$-wave heavy baryons of the $SU(3)$ flavor $\mathbf{6}_F$,''
  Phys.\ Rev.\ D {\bf 96}, no. 7, 074021 (2017).


%
\bibitem{Wang:2017xam}
  Z.~G.~Wang, X.~N.~Wei and Z.~H.~Yan,
  ``Revisit assignments of the new excited $\Omega_c$ states with QCD sum rules,''
  Eur.\ Phys.\ J.\ C {\bf 77}, no. 12, 832 (2017).





%
\bibitem{OmegacLattice}
  M.~Padmanath and N.~Mathur,
  ``Quantum Numbers of Recently Discovered $\Omega^{0}_{c}$ Baryons from Lattice QCD,''
  Phys.\ Rev.\ Lett.\  {\bf 119}, no. 4, 042001 (2017).



%
\bibitem{yang}
  G.~Yang and J.~Ping,
  ``The structure of pentaquarks $\Omega_c^0$ in the chiral quark model,''
  arXiv:1703.08845 [hep-ph].

%
\bibitem{huangping}
  H.~Huang, J.~Ping and F.~Wang,
  ``Investigating the excited $\Omega^{0}_{c}$ states through $\Xi_{c}K$ and $\Xi^{'}_{c}K$ decay channels,''
  arXiv:1704.01421 [hep-ph].

%
\bibitem{polyakov}
  H.~C.~Kim, M.~V.~Polyakov and M.~Prasza{\l}owicz,
  ``Possibility of the existence of charmed exotica,''
  Phys.\ Rev.\ D {\bf 96}, no. 1, 014009 (2017);
  Addendum: [Phys.\ Rev.\ D {\bf 96}, no. 3, 039902 (2017)].

%
\bibitem{an}
  C.~S.~An and H.~Chen,
  ``Observed $\Omega_{c}^{0}$ resonances as pentaquark states,''
  Phys.\ Rev.\ D {\bf 96}, no. 3, 034012 (2017).

%
\bibitem{sheldon}
  A.~Ali, J.~S.~Lange and S.~Stone,
  Exotics: Heavy Pentaquarks and Tetraquarks,''
  Prog.\ Part.\ Nucl.\ Phys.\  {\bf 97}, 123 (2017).

%
\bibitem{anisovich}
  V.~V.~Anisovich, M.~A.~Matveev, J.~Nyiri and A.~N.~Semenova,
  ``Narrow pentaquarks as diquark-diquark-antiquark systems,''
  Mod.\ Phys.\ Lett.\ A {\bf 32}, no. 29, 1750154 (2017).



%
\bibitem{zezhao}
  Z.~Zhao, D.~D.~Ye and A.~Zhang,
  ``Hadronic decay properties of newly observed $\Omega_c$ baryons,''
  Phys.\ Rev.\ D {\bf 95}, no. 11, 114024 (2017).

%
\bibitem{wangzhao}
  K.~L.~Wang, Y.~X.~Yao, X.~H.~Zhong and Q.~Zhao,
  ``Strong and radiative decays of the low-lying $S$- and $P$-wave singly heavy baryons,''
  Phys.\ Rev.\ D {\bf 96}, no. 11, 116016 (2017).

%
\bibitem{kim}
  H.~C.~Kim, M.~V.~Polyakov, M.~Prasza{\l}owicz and G.~S.~Yang,
  ``Strong decays of exotic and non-exotic heavy baryons in the chiral quark-soliton model,''
  Phys.\ Rev.\ D {\bf 96}, no. 9, 094021 (2017).

%
\bibitem{chengchiang}
  H.~Y.~Cheng and C.~W.~Chiang,
  ``Quantum numbers of $\Omega_c$ states and other charmed baryons,''
  Phys.\ Rev.\ D {\bf 95}, no. 9, 094018 (2017).








%
\bibitem{hofmann}
  J.~Hofmann and M.~F.~M.~Lutz,
  ``Coupled-channel study of crypto-exotic baryons with charm,''
  Nucl.\ Phys.\ A {\bf 763}, 90 (2005).

%
\bibitem{tejero}
  C.~E.~Jimenez-Tejero, A.~Ramos and I.~Vidana,
  ``Dynamically generated open charmed baryons beyond the zero range approximation,''
  Phys.\ Rev.\ C {\bf 80}, 055206 (2009).

%
\bibitem{romanets}
  O.~Romanets, L.~Tolos, C.~Garcia-Recio, J.~Nieves, L.~L.~Salcedo and R.~G.~E.~Timmermans,
  ``Charmed and strange baryon resonances with heavy-quark spin symmetry,''
  Phys.\ Rev.\ D {\bf 85}, 114032 (2012).




%
\bibitem{OmegacPavao}
  J.~Nieves, R.~Pavao and L.~Tolos,
  ``$\Omega_c$ excited states within a ${\rm SU(6)}_{\rm lsf}\times$HQSS model,''
  arXiv:1712.00327 [hep-ph].

%
\bibitem{OmegacMontana}
  G.~Montana, A.~Feijoo and A.~Ramos,
  ``A meson-baryon molecular interpretation for some $\Omega_c$ excited baryons,''
  arXiv:1709.08737 [hep-ph].

%
\bibitem{OmegacDebastiani}
  V.~R.~Debastiani, J.~M.~Dias, W.~H.~Liang and E.~Oset,
  ``Molecular $\Omega_c$ states generated from coupled meson-baryon channels,''
  arXiv:1710.04231 [hep-ph].





\bibitem{hidden1}
  M.~Bando, T.~Kugo, S.~Uehara, K.~Yamawaki and T.~Yanagida,
  ``Is rho Meson a Dynamical Gauge Boson of Hidden Local Symmetry?,''
  Phys.\ Rev.\ Lett.\  {\bf 54}, 1215 (1985).

\bibitem{hidden2}
  M.~Bando, T.~Kugo and K.~Yamawaki,
  ``Nonlinear Realization and Hidden Local Symmetries,''
  Phys.\ Rept.\  {\bf 164}, 217 (1988).

\bibitem{hidden4}
  U.~G.~Mei{\ss}ner,
  ``Low-Energy Hadron Physics from Effective Chiral Lagrangians with Vector Mesons,''
  Phys.\ Rept.\  {\bf 161}, 213 (1988).

\bibitem{Yama}
  M.~Harada and K.~Yamawaki,
  ``Hidden local symmetry at loop: A New perspective of composite gauge boson and chiral phase transition,''
  Phys.\ Rept.\  {\bf 381}, 1 (2003).

\bibitem{hideko}
  H.~Nagahiro, L.~Roca, A.~Hosaka and E.~Oset,
  ``Hidden gauge formalism for the radiative decays of axial-vector mesons,''
  Phys.\ Rev.\ D {\bf 79}, 014015 (2009).




\bibitem{isgur}
  N.~Isgur and M.~B.~Wise,
  ``Weak Decays of Heavy Mesons in the Static Quark Approximation,''
  Phys.\ Lett.\ B {\bf 232}, 113 (1989).

\bibitem{neubert}
 M.~Neubert,
  ``Heavy-Quark Symmetry''
  Phys.\ Rept.\  {\bf 245}, 259 (1994).

\bibitem{manohar}
  A.~V.~Manohar and M.~B.~Wise,
  ``Heavy quark physics,''
  Camb.\ Monogr.\ Part.\ Phys.\ Nucl.\ Phys.\ Cosmol.\  {\bf 10}, 1 (2000).










%
\bibitem{Omegac3188}
  C.~Wang, L.~L.~Liu, X.~W.~Kang and X.~H.~Guo,
  ``Possible open-charmed pentaquark molecule $\Omega_c(3188)$ --- the $D \Xi$ bound state --- in the Bethe-Salpeter formalism,''
  arXiv:1710.10850 [hep-ph].

%
\bibitem{OmegacHosaka}
  R.~Chen, X.~Liu and A.~Hosaka,
  ``Searching for possible $\Omega_c-$like molecular states from $\Xi_{c}^{*}\bar{K}/\Xi_c\bar{K}^*/\Xi_c'\bar{K}^*$ interaction,''
  arXiv:1711.07650 [hep-ph].

%
\bibitem{OmegacGeng}
  Y.~Huang, C.~j.~Xiao, Q.~F.~L\"{u}, R.~Wang, J.~He and L.~Geng,
  ``Strong and radiative decays of $D\Xi$ molecular state and newly observed $\Omega_c$ states,''
  arXiv:1801.03598 [hep-ph].







%
\bibitem{Belyaev}
I. Belyaev, ``Spectroscopy of charm baryons at LHCb'', talk presented at the Hadron 2017 Conference, Salamanca, September 2017,
\href{https://indico.cern.ch/event/578804/contributions/2636994/}{https://indico.cern.ch/event/578804/contributions/2636994/}



%
\bibitem{OmegabLHCb}
  R.~Aaij {\it et al.} [LHCb Collaboration],
  ``Measurement of the mass and lifetime of the $\Omega_b^-$ baryon,''
  Phys.\ Rev.\ D {\bf 93}, no. 9, 092007 (2016).

%
\bibitem{OmegabLHCbprevious}
  R.~Aaij {\it et al.} [LHCb Collaboration],
  ``Measurement of the $\Lambda_b^0$, $\Xi_b^-$ and $\Omega_b^-$ baryon masses,''
  Phys.\ Rev.\ Lett.\  {\bf 110}, no. 18, 182001 (2013).

%
\bibitem{OmegabCDF}
  T.~A.~Aaltonen {\it et al.} [CDF Collaboration],
  ``Mass and lifetime measurements of bottom and charm baryons in $p\bar p$ collisions at $\sqrt{s}= 1.96$ TeV,''
  Phys.\ Rev.\ D {\bf 89}, no. 7, 072014 (2014).

%
\bibitem{OmegabD0}
  V.~M.~Abazov {\it et al.} [D0 Collaboration],
  ``Observation of the doubly strange $b$ baryon $\Omega_b^-$,''
  Phys.\ Rev.\ Lett.\  {\bf 101}, 232002 (2008).

%
\bibitem{pdg2016}
C.~Patrignani {\it et al.} (Particle Data Group),
 ``Review of Particle Physics,'' (with 2017 update),
 Chin.\ Phys.\ C {\bf 40}, 100001 (2016).


\bibitem{nsd}
  J.~A.~Oller and E.~Oset,
  ``N/D description of two meson amplitudes and chiral symmetry,''
  Phys.\ Rev.\ D {\bf 60}, 074023 (1999).

\bibitem{ollerulf}
  J.~A.~Oller and U.~G.~Mei{\ss}ner,
  ``Chiral dynamics in the presence of bound states: Kaon nucleon interactions revisited,''
  Phys.\ Lett.\ B {\bf 500}, 263 (2001).

\bibitem{gamermann}
  D.~Gamermann, J.~Nieves, E.~Oset and E.~Ruiz Arriola,
  ``Couplings in coupled channels versus wave functions: application to the X(3872) resonance,''
  Phys.\ Rev.\ D {\bf 81}, 014029 (2010).

\bibitem{acetifun}
  F.~Aceti and E.~Oset,
  ``Wave functions of composite hadron states and relationship to couplings of scattering amplitudes for general partial waves,''
  Phys.\ Rev.\ D {\bf 86}, 014012 (2012).

\bibitem{lianghadro}
  W.~H.~Liang, M.~Bayar and E.~Oset,
  ``$\Lambda_b \to \pi^- (D_s^-) \Lambda_c(2595),~\pi^- (D_s^-) \Lambda_c(2625)$ decays and $DN,~D^*N$ molecular components,''
  Eur.\ Phys.\ J.\ C {\bf 77}, no. 1, 39 (2017).

\bibitem{bramon}
  A.~Bramon, A.~Grau and G.~Pancheri,
  ``Intermediate vector meson contributions to V0 $\to$ P0 P0 gamma decays,''
  Phys.\ Lett.\ B {\bf 283}, 416 (1992).

\bibitem{Close}
  F.~E.~Close,
  ``An Introduction to Quarks and Partons,''
  Academic Press/London 1979, 481p.

\bibitem{miyahara}
  K.~Miyahara, T.~Hyodo, M.~Oka, J.~Nieves and E.~Oset,
  ``Theoretical study of the $\Xi(1620)$ and $\Xi(1690)$ resonances in $\Xi_c \to \pi^+ MB$ decays,''
  Phys.\ Rev.\ C {\bf 95}, no. 3, 035212 (2017).

\bibitem{Pavao}
  R.~P.~Pavao, W.~H.~Liang, J.~Nieves and E.~Oset,
  ``Predictions for $\Xi_b^- \rightarrow \pi^- \left(D_s^- \right) \ \Xi_c^0 (2790) \left(\Xi_c^0 (2815) \right)$ and $\Xi_b^- \rightarrow \bar{\nu}_l l \ \Xi_c^0 (2790) \left(\Xi_c^0 (2815) \right)$,''
  Eur.\ Phys.\ J.\ C {\bf 77}, no. 4, 265 (2017).

\bibitem{mizutani}
  T.~Mizutani and A.~Ramos,
  ``D mesons in nuclear matter: A DN coupled-channel equations approach,''
  Phys.\ Rev.\ C {\bf 74}, 065201 (2006).

%
%
%
%
%













%
%
%
%
%
%
%
%
%
%
%
%
%
%
%
%
%
%
%
%
%
%
%
%
%
%
%
%
%
%
%
%
%
%
%
%
%
%
%
%





\end{thebibliography}
  \end{document}